\documentclass[epj,aps,twocolumn,preprintnumbers, showpacs, nofootinbib,superscriptaddress,notitlepage]{revtex4-1}
\usepackage{amsmath,graphicx,epsfig,amssymb}
\usepackage{inputenc}
\usepackage[usenames]{color}
\usepackage{ulem} 
\usepackage{bigstrut}
\usepackage{slashed}
\usepackage{multirow}
\usepackage{subfigure}
\usepackage{dsfont}
\usepackage{xcolor}
\usepackage{hyperref}
\usepackage{tabularx} 
\usepackage{booktabs}

\begin{document}
\title{Investigate the glueball-like particle $X(2370)$ in $B$ meson decays}
\author{Xiao-Tong Li}
\affiliation{Shanghai Key Laboratory for Particle Physics and Cosmology, Key Laboratory for Particle Astrophysics and Cosmology (MOE), School of Physics and Astronomy, Shanghai Jiao Tong University, Shanghai 200240, China}

\author{Guang-Yu Wang}
\email{Corresponding author: guangyuwang@sjtu.edu.cn}
\affiliation{Shanghai Key Laboratory for Particle Physics and Cosmology, Key Laboratory for Particle Astrophysics and Cosmology (MOE), School of Physics and Astronomy, Shanghai Jiao Tong University, Shanghai 200240, China}
  
\author{Qi-An Zhang} 
\email{Corresponding author: zhangqa@buaa.edu.cn}
\affiliation{School of Physics, Beihang University, Beijing 102206, China}

\begin{abstract}
Based on the collected data on  $J/\psi\to \gamma K_S^0K_S^0\eta'$,  the BESIII experiment has conducted an analysis of the mass and spin parity of the $X(2370)$ particle. The findings are consistent with the characteristics expected of the lightest pseudoscalar glueball. We point out that further exploration of this particle’s nature can be pursued through investigations of heavy bottom meson decays. Assuming the identity of $X(2370)$ as a pseudoscalar glueball, we compute the form factors for $B\to X(2370)$ transitions in the factorization approach. With these results,  the estimated branching fractions for semileptonic $B$  decays into $X(2370)$ can reach  the order of $10^{-6}$ and those for nonleptonic decays can reach the order $10^{-8}$.  These results suggest that decays of $B$ meson into $X(2370)$  are detectable at   experimental facilities like Belle-II. Future experimental endeavors hold promise in expanding our understanding of glueball physics, contributing to the ongoing exploration  surrounding this intriguing particle.
\end{abstract}

\maketitle

\section{Introduction}

The existence of glueballs is predicted by QCD, the theory that describes the strong interactions and strong nuclear force. Observing and studying glueballs would provide experimental confirmation of QCD and  can deepen our understanding of how quarks and gluons interact to form the particles we see in the universe. Thereby there are a number of experimental measurements which have found different candidates for glueballs (for reviews please see Refs.~\cite{Klempt:2007cp,Crede:2008vw}).  Very recently based on the available  events collected with the BESIII detector, a partial wave analysis of the decay  $J/\psi\to \gamma K_S^0K_S^0\eta'$ has been  performed in Ref.~\cite{BESIII:2023wfi}.  In this process,  the mass and width of  $X(2370)$ discovered in Ref.~\cite{Liu:2010tr} are determined, and it is found that  the measured mass and spin parity  are consistent with the predictions of a  lightest pseudoscalar glueball~\cite{Gui:2019dtm}.   With a spin-parity configuration of $0^{-+}$, this  particle challenges conventional models and infers a reevaluation of existing theoretical frameworks. Its property hints at a rich spectrum of exotic states that remain to be uncovered, promising a wealth of new phenomena to be unraveled.

In the literature, there have been tremendous studies on the glueball candidates from lattice QCD~\cite{Bali:1993fb,Morningstar:1999rf,Chen:2005mg,Gregory:2012hu}. Recently the  Lattice QCD explorations are focused  on the spectrum of scalar glueball~\cite{Gui:2012gx,Zou:2024ksc}, tensor glueball~\cite{Yang:2013xba},  pseudoscalar~\cite{Gui:2019dtm} and $\eta$-glueball mixing~\cite{Jiang:2022ffl}.  In particular Ref.~\cite{Gui:2019dtm} has calculated the  production rate of the pseudoscalar glueball in  $J/\psi$ radiative decays, and these results are in the right ballpark with the BESIII measurement~\cite{BESIII:2023wfi}, which leads to the conjecture that the $X(2370)$ might be a pseudoscalar glueball.

The deciphering of the internal structure and mysterious properties  of the $X(2370)$ can proceed not only through the detailed analysis of the mass and decay width, but also through the decay and production characters~\cite{Yu:2011ta,Deng:2012wi,She:2024ewy,Cao:2024mfn,Li:2024fko}.   In this work, we will  point out that in addition to $J/\psi$ radiative decays which are regarded as an important hunting ground for glueballs, owing to its the gluonrich environment and clean background,  semileptonic and nonleptonic decays of  heavy bottom meson  can also be a platform to investigate the glueball interpretation for $X(2370)$.  Following Refs.~\cite{He:2006qk,Charng:2006zj,Wang:2009rc,Lu:2013jj,Huang:2021ots,Wang:2017hxe,Zhou:2016jkv} we will delve into the implications by investigating the properties of X(2370) within the context of $B$ meson decays. We will  point out that the semileptonic $\overline B^0\to X(2370)\ell\bar\nu$ and $B^\pm \to X(2370)\pi^\pm /K^\pm$ can be used to validate the existence of this particle and explore its glueball nature. By making use of the factorization scheme~\cite{Keum:2000wi,Keum:2000ph,Lu:2000em}, we present an estimate of the transition form factors, which are subsequently used to determine the  corresponding decay branching fractions. Our findings suggest that the decay branching fraction of $B\to X(2370)$ decays   lies at the order of $10^{-6}$ to $10^{-8}$, indicating a rare but observable phenomenon at experimental facilities such as Belle-II~\cite{Belle-II:2018jsg}. Future experimental investigation could serve as a signature for identifying and probing this exotic particle.

The subsequent sections of this paper are structured as follows. Sec.~\ref{sec:factorization} presents the theoretical framework utilized for computing the transition form factors under the factorization framework. These results are then employed to determine the decay branching fractions for semileptonic decays of $B$ mesons and nonleptonic section in Sec.~\ref{sec:pheno}. A brief summary is given in the last section. The appendix collects some necessary details in the calculation.

\section{Transition form factors}
\label{sec:factorization}

\begin{figure}
\begin{center}
\includegraphics[scale=0.5]{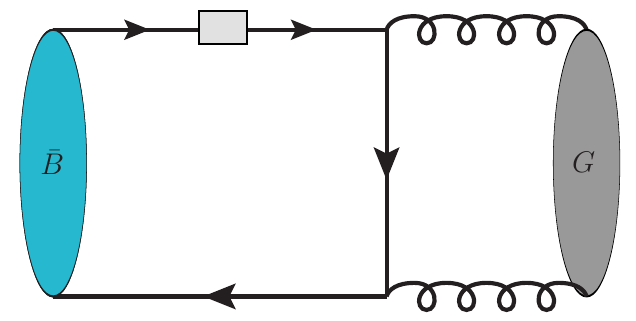}
\caption{Feynman diagrams of $\overline{B}$ decays into a pseudo-scalar glueball
$G$. The grey box denotes the possible Lorentz structure arising from the
electroweak interactions.} 
\label{diagram:FeynBtof0}
\end{center}
\end{figure}

As an estimate we will adopt the perturbative QCD (PQCD) approach~\cite{Keum:2000wi,Keum:2000ph,Lu:2000em} based on $k_T$ factorization to compute the transition form factor in which the leading-order Feynman diagram for the $B\to G l\bar\nu$ decays is displayed in Fig~\ref{diagram:FeynBtof0}.  Other Feynman diagrams are power suppressed as pointed out in Ref.~\cite{Wang:2009rc}.  We examine the kinematics of these decays within the large-recoil (low $q^2$) regime, where the PQCD factorization method is deemed applicable for the semileptonic decays under consideration. In the rest frame of the $B$ meson, we define the $B$ meson momentum $P_{B}$ and the final glueball momentum $P_G$ in the light-cone coordinates:
 \begin{eqnarray}
 P_{B}=\frac{m_{B}}{\sqrt{2}}(1,1,0_{\perp})\;,\;
 P_G=\frac{m_{B}}{\sqrt{2}}(\rho,0,0_{\perp})\;.\label{eq:momentum}
 \end{eqnarray}
The energy fraction $\rho$ is approximately $\rho\approx1-\frac{q^2}{m_{B}^2}$, with $q=P_{B}-P_G$ denoting the momentum transfer to the lepton pair.

\subsection{Lightcone wave functions}

The $B$ meson light-cone matrix element has been
given in Refs.~\cite{Grozin:1996pq,Kawamura:2001jm}, which can be decomposed into the form as
 \begin{eqnarray}
 &&\int_0^1\frac{d^4z}{(2\pi)^4}e^{ik_1\cdot z}\langle 0|b_{\beta}(0)\bar
 q_{\alpha}(z)|\bar B(P_{B})\rangle \nonumber\\
 && =
 \frac{i}{\sqrt{2N_c}}\left\{(\not\! P_{B}+m_{B})\gamma_5\left[\phi_{B}(k_1)+\frac{\not\!
 n }{\sqrt{2}}\bar\phi_{B}(k_1)\right]\right\}_{\beta\alpha}\;,\label{Bwav:decompose}
 \end{eqnarray}
with $n=(1,0,\textbf{0}_T)$ and $v=(0,1,\textbf{0}_T)$ are light-like unit vectors. In this paper, we focus solely on the contribution from 
$\phi_{B}(k_1)$, as the impact of $\bar\phi_{B}(k_1)$ is found to be quantitatively minor and is thus neglected. 

For the $B$-meson wave function, we adopt the form widely used in the PQCD approach \cite{Ball:2004ye,Hu:2012cp} as follows:
\begin{eqnarray}
\phi_{B}(x,b)=N_{B}x^2(1-x)^2\mbox{exp}\left[-\frac{m_{B}^2 x^2}{2\omega_b^2}-\frac{1}{2}(\omega_b
b)^2\right], \label{Bwave:da}
 \end{eqnarray}
 which has been used in tremendous phenomenological studies. Notice that this parametrization is model-dependent and has been tuned in the PQCD approach. 
 The normalization factors $N_{B}$ is chosen to be connected with the decay constants $f_{B}$ by the following relationship:
\begin{eqnarray}
\int\frac{d^4 k_1}{(2\pi)^4}\phi_{B}({
k_1})=\frac{f_{B}}{2\sqrt{2N_c}}\;,\; \int \frac{d^4
k_1}{(2\pi)^4}\bar{\phi}_{B}({ k_1})=0.\label{Bwave:normalization}
\end{eqnarray}
In this parametrization~\cite{Keum:2000wi,Keum:2000ph,Lu:2000em}, one usually  sets the shape parameter $\omega_b$ to $0.40  \mathrm{GeV}$, informed by extensive experimental data on $B$ mesons \cite{Li:2002uya}, resulting in a normalization constant $N_B$ equal of $91.736$ when $f_{B}=0.19 \mathrm{GeV}$~\cite{ParticleDataGroup:2024cfk}. It is important to note that there are lattice QCD calculations of the $B$ meson LCDAs~\cite{Han:2024min,Han:2024yun,Wang:2024wwa}  where the transverse separation is not included. A direct calculation of $B$ meson wave functions, including the transverse distributions, similar to those for light mesons and baryons \cite{LatticeParton:2020uhz,LatticePartonLPC:2022eev, LatticeParton:2023xdl,LatticePartonLPC:2023pdv,Chu:2024vkn,LatticeParton:2022zqc,Liu:2018tox,Liu:2019urm,LatticeParton:2018gjr,Hua:2020gnw,Xu:2018mpf,LatticePartonCollaborationLPC:2022myp}, may become possible in future.

In general, defining the LCDAs for a glueball requires the gauge invariant building block using field strength tensor.
In the $A^+=0$ gauge, the light cone distribution amplitude of a pseudo-scalar glueball state is defined as:
\begin{eqnarray}
&&\left\langle G(P)\right| A_{[\mu}^{a}(z) A_{\nu]}^{b}(0)|0\rangle \nonumber\\
&&= \frac{f_{s}}{2} \frac{\delta^{a b}}{N_{c}^{2}-1} \epsilon_{\mu \nu \rho \sigma} \frac{n_{-}^{\rho} P^{\sigma}}{n_{-} \cdot P} \int_{0}^{1} d x e^{i x P \cdot z} \frac{\phi_{G}(x)}{x(1-x)}
,\label{eq:definition-gauge-invariant-LCDA}
\end{eqnarray}
where the notation $A_{[\mu}^a(z)A_{\nu]}^b(0)\equiv[A_{\mu}^a(z)A_{\nu}^b(0)-A_{\nu}^a(z)A_{\mu}^b(0)]/2$, and the estimate of decay constant $f_s=0.13 \mathrm{GeV}$ \cite{He:2002hr}. Due to the limited research available on the pseudoscalar glueball distribution amplitude, our previous work primarily relied on the distribution amplitudes of pseudoscalar mesons as described, in Ref.\cite{Charng:2006zj,Ali:2003kg} The normalization factor for the distribution amplitude of a pseudoscalar glueball is referenced to that of the scalar glueball's distribution amplitude\cite{Wang:2009rc}. The function $\phi_G(x)$ is expressed as:
\begin{eqnarray}
\phi_G(x)&=&
30x^2(1-x)^2\left[1+\sum_{n}a_n C_n^{5/2}(2x-1)\right],\label{eq:LCDAgegenbauermoments}
\end{eqnarray}
where the coefficient $a_2=0.2$ for the Gegenbauer moment \cite{Wang:2009rc} can be adopted as an estimate.  This value is used in the absence of theoretical studies on the LCDAs of pseudoscalar glueballs, with the simplest approach being the application of the asymptotic form. This choice of $a_2=0.2$ allows us to roughly estimate the uncertainty associated with the higher Gegenbauer moments in our analysis. $C_n^{5/2}(t)$ denotes the Gegenbauer polynomial.

\subsection{Transition form factors}
\label{sec3}

For the $B \to M$ transition with $M$ being a pseudoscalar meson, the pertinent form factors $F_1 (q^2)$ and $F_0 (q^2)$, as well as $F_T (q^2)$, have been defined~\cite{Wang:2009rc}, with the condition $F_1 (0)=F_0 (0)$.The form factors for $B \to G$ are defined via
\begin{widetext}
\begin{eqnarray}
\langle G(P_G) |\bar q \gamma_\mu b|\bar {B}(P_{B})\rangle
&=&F_{1}(q^2)\left[(P_B+P_G)_\mu
-\frac{m_B^2-m_{G}^2}{q^2}q_\mu\right]+F_{0}(q^2)
\frac{m_B^2-m_{G}^2}{q^2}q_\mu\;,\nonumber\\
\langle G(P_G)|\bar q  i\sigma^{\mu\nu} q_\nu
b|\bar {B}(P_{B}) \rangle &=& \frac{
F_T(q^2)}{m_B+m_{G}}\left[
(m_B^2-m_{G}^2)\,q^\mu-q^2(P_B^\mu+P_G^{\mu})\right]\;.
\label{ftensor}
\end{eqnarray}
\end{widetext}

In the hard-scattering kernel, the transverse momentum terms in the denominators are usually retained to regulate the endpoint singularities and in the present case the results are not sensitive to the transverse momentum due to the absence of endpoint singularity. The contributions proportional to $k_{1T}$ and $k_{2T}$ in the numerator are discarded as they are power-suppressed relative to the other terms. Incorporating the Sudakov form factors and threshold resummation effects in the transverse configuration $b$-space, we derive the $B \to G$  form factors as follows:
\begin{widetext}
\begin{eqnarray}
F_{1}(q^2)&=&8\sqrt{6}\pi
m_B^2 f_s\frac{C_F}{N_C^2-1}\int dx_1dx_2\int b_1db_1
b_2db_2\phi_B(x_1,b_1)\frac{\phi_G(x_2)}{x_2(1-x_2)}
\nonumber\\
& &\times x_1[2+(\rho-1)x_2-\rho]E(t)h(x_1,x_2,b_1,b_2)\;,\nonumber\\
F_{0}(q^2)&=&8\sqrt{6}\pi
m_B^2 f_s\frac{C_F}{N_C^2-1}\int dx_1dx_2\int b_1db_1
b_2db_2\phi_B(x_1,b_1)\frac{\phi_G(x_2)}{x_2(1-x_2)}
\nonumber \\
& &\times x_1\rho[(1-\rho)x_2+\rho]E(t)h(x_1,x_2,b_1,b_2)\;,\nonumber\\
F_{T}(q^2)&=&8\sqrt{6}\pi
m_B^2 f_s\frac{C_F}{N_C^2-1}\int dx_1dx_2\int b_1db_1
b_2db_2\phi_B(x_1,b_1)\frac{\phi_G(x_2)}{x_2(1-x_2)}
\nonumber\\
& &\times x_1(2-x_2)E(t)h(x_1,x_2,b_1,b_2)\;, \label{gfpi}
\end{eqnarray}
\end{widetext}
with the color factor $C_F=\frac{4}{3}$ and $N_C=3$. The impact parameters $b_{1}$ and $b_{2}$ conjugate to $k_{1T}$ and $k_{2T}$, respectively, in conjunction with the hard function $h(x_1, x_2, b_1, b_2)$ and the evolution factor\cite{Li:2001ay,Lim:1998uc} 
\begin{eqnarray}
E(t) = \alpha_s(t)e^{-S_B(t)-S_{G}(t)}. \label{evol}
\end{eqnarray}
The hard function and the Sudakov factors ($S_B(t)$, $S_{G}(t)$), detailed in the Appendix, are included. In the process of evaluating the Sudakov factors mentioned previously, we incorporate the one-loop expression for the running coupling constant, denoted as $\alpha_s(t)$.  By adopting the one-loop approximation, we are able to capture the essential features of the coupling constant’s behavior as a function of the energy scale. The hard scales $t$ in the equations of this work are chosen as 
\begin{eqnarray}
t= {\rm max}\left[\sqrt {\rho x_1}m_B,1/b_1,1/b_2\right],
\end{eqnarray}
which represent the largest scale of the virtuality of the internal particles in the hard b-quark decay diagram.

The input parameters listed above are crucial for the accuracy of our numerical computations. The choice of $\Lambda_{\overline{MS}}^{(f=4)}=0.25 \mathrm{GeV}$ is based on previous studies that align with our computational framework. The masses of the $B$ meson, $m_B$ is taken as  $5.279 \mathrm{GeV}$, respectively, reflecting the latest experimental results~\cite{ParticleDataGroup:2024cfk}. 
The mass of the glueball is taken as $m_G=2.4 \mathrm{GeV}$, in Ref.\cite{BESIII:2023wfi}. 
In our analysis, we focus on the PQCD predictions for the form factors $F_{1,0,T}(q^2)$ within the region where they are considered reliable $0 \leq q^2 \leq 10 \mathrm{GeV}^2$. To extend our predictions to larger values of $q^2$, we first employ a pole model parametrization given by 
\begin{eqnarray}
F_i(q^2)=\frac{F_i(0)}{1-a(q^2/m_B^2)+b(q^2/m_B^2)^2},\label{eq:fitting-form}
\end{eqnarray}
where $F_i$ denotes a function among $F_{1,0,T}$, and $a, b$ are
the constants to be determined by the fitting procedure. This parametrization allows us to extrapolate the form factors beyond the region of direct PQCD calculation, certifying that our results are applicable over a broader range of $q^2$ values.

\begin{table}
\caption{Distinct contributions to $B\to G$ form factors at $q^2=0$. The difference in the form factor values for $F_T(0)$ and $F_1(0)$ is evident, with $F_T(0)$ including an additional term proportional to $x_2$. Furthermore, $F_1(0)$ and $F_0(0)$ are identical when $q^2=0$.}\label{tab:formfactor}
\renewcommand{\arraystretch}{2.0}
\setlength{\tabcolsep}{2.5mm}
\begin{tabular}{c|c|c|c}
\hline\hline & $F_i(0)$ & $a$& $b$   \\ \hline
 $F_1$  &$0.088^{+0.013+0.008}_{-0.013-0.010}$& $1.79^{+0.04}_{-0.02}$ & $0.48^{+0.08}_{-0.07}$ \\ \hline
 $F_0$  &$0.088^{+0.013+0.008}_{-0.013-0.010}$& $0.27^{+0.04}_{-0.07}$  & $-0.69^{+0.10}_{-0.09}$ \\ \hline
 $F_T$  &$0.114^{+0.023+0.021}_{-0.016-0.005}$ & $1.65^{+0.02}_{-0.02}$ & $0.42^{+0.08}_{-0.07}$   \\
\hline\hline
\end{tabular}
\end{table}

For comparative purposes, we also use $z$-series parametrizations to extend the PQCD calculations to high $q^2$ regions. In this work, we adopt the Bourrely-Caprini-Lellouch (BCL) version of the $z$-series expansion  \cite{Bourrely:2008za,Zhang:2021oja}.
In this approach, the form factor can be expanded as
\begin{eqnarray}
&&F_{B \to G}^{i}(q^2) =
{F_{B \to G}^{i}(0) \over 1-q^2/m_{i, \, \rm pole}^2} \nonumber\\
&&
\times \left \{1 + \sum_{k=1}^N \, b_k^i \, \left [z(q^2, t_0)^k - z(0, t_0)^k \right ]  \right \}, 
\label{z-series expansion}
\end{eqnarray}
where 
\begin{eqnarray}
z(q^2, t_0) = \frac{\sqrt{t_{+}-q^2}-\sqrt{t_{+}-t_0}}{\sqrt{t_{+}-q^2}+\sqrt{t_{+}-t_0}}\,,
\end{eqnarray}
with $t_{+}=(m_B +m_G)^2$ and $t_0=(m_B + m_G)(\sqrt{m_B}-\sqrt{m_G})^2$. For $i=1,T$, the pole mass is taken as $m_{i, pole}=m_{B^*(1-)}=5.324\rm{GeV}$ while for $i=0$, $m_{i, pole}=m_{B^*(0+)}=5.627\rm{GeV}$~\cite{ParticleDataGroup:2024cfk}.

\begin{table}
\caption{The form factors for the $B\to G$ transition are calculated using the $z$-series expansion within the framework of PQCD factorization.}\label{tab:formfactor2}
\renewcommand{\arraystretch}{2.0}
\setlength{\tabcolsep}{2.5mm}
\begin{tabular}{c|c|c}
\hline\hline & $F_i(0)$ & $b_1$   \\ \hline
 $F_1$  &$0.083^{+0.021+0.010}_{-0.010-0.005}$& $-12.20^{+0.07}_{-0.07}$\\ \hline
 $F_0$  &$0.083^{+0.021+0.010}_{-0.010-0.005}$& $3.68^{+0.10}_{-0.09}$\\ \hline
 $F_T$  &$0.112^{+0.026+0.013}_{-0.013-0.008}$ & $-8.02^{+0.07}_{-0.07}$\\
\hline\hline
\end{tabular}
\end{table}

The predictions for the form factors $F_1$,$F_0$ and $F_T$ (at $q^2 = 0$) in the $B\to G$ transitions are presented in table~\ref{tab:formfactor} and table~\ref{tab:formfactor2}. The sources of the first two errors are from  those for the $B$ meson wave function: 
$\omega_B=(0.40\pm0.05)\mbox{GeV}$ and $f_B=(0.19\pm0.02)\rm{GeV}$
for $B$ mesons \cite{Keum:2000ph}. More uncertainties can be added but for the present estimate these two kinds of errors should be enough. 

\begin{figure}
\begin{center}
\includegraphics[scale=0.6]{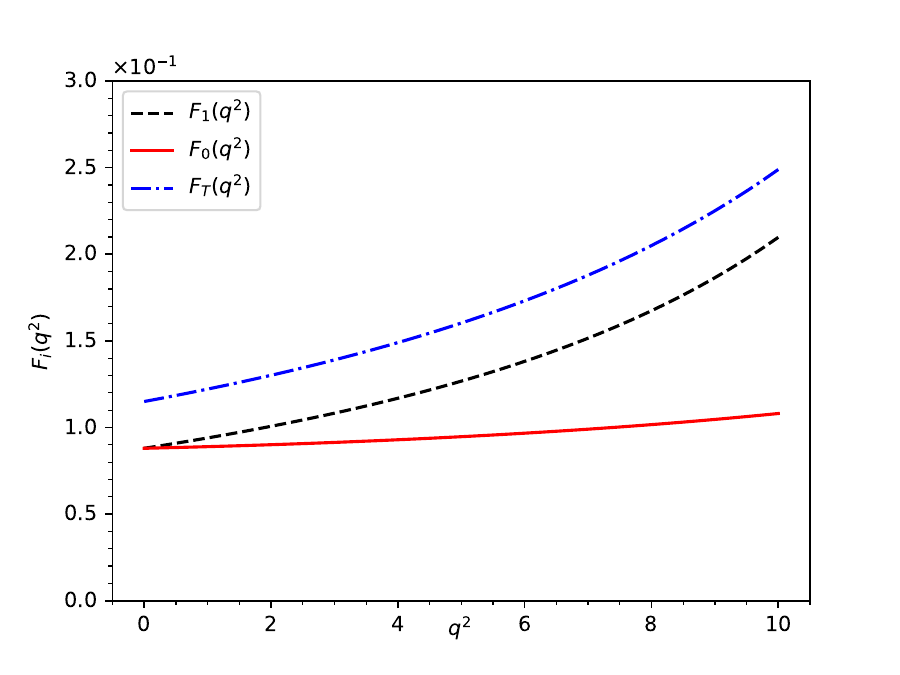}
\hspace{1cm}
\includegraphics[scale=0.6]{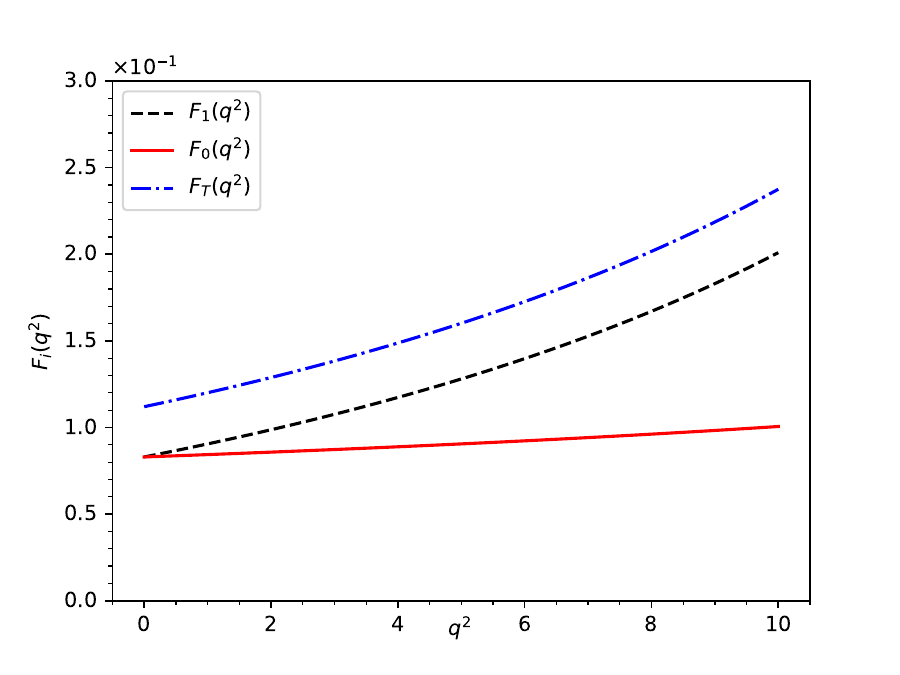}
\end{center}
\caption{The $q^2$-dependence of $B \to G$ form factors
$|F_1(q^2)|$, $|F_0(q^2)|$ and $|F_T(q^2)|$ using the dipole parametrization (upper panel) and $z$-series expansion (lower
panel) within the framework of PQCD factorization.} \label{diagram:q2dependence}
\end{figure}

\section{PHENOMENOLOGICAL APPLICATIONS IN DECAY}
\label{sec:pheno}

\subsection{Semileptonic $B\rightarrow Gl\Bar{\nu}$ and $B\rightarrow G\tau\Bar{\nu_\tau}$ Decays}

\begin{figure}
\begin{center}

\includegraphics[scale=0.6]{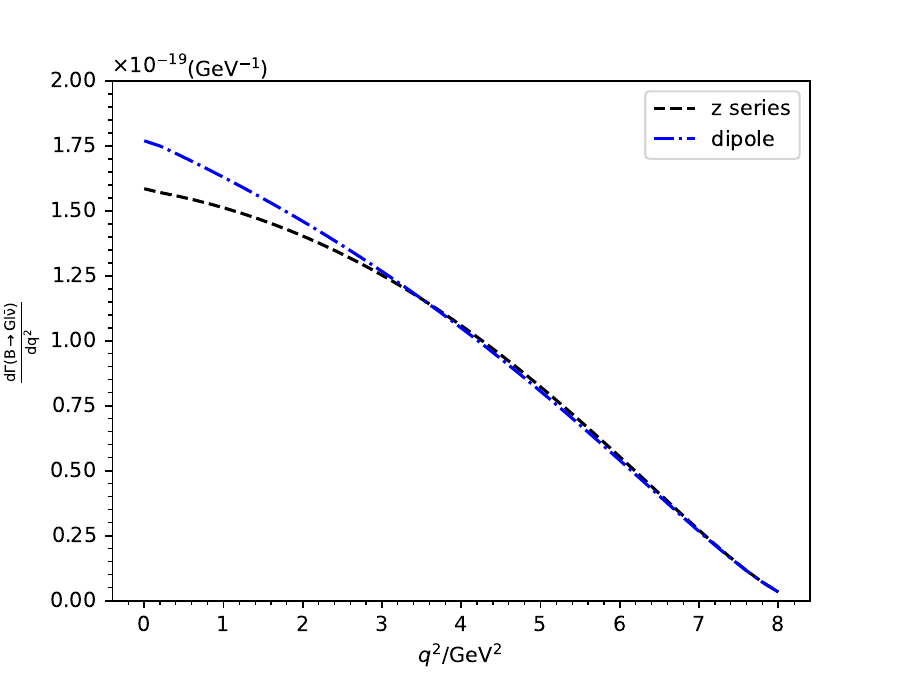}\hspace{1cm}
\includegraphics[scale=0.6]{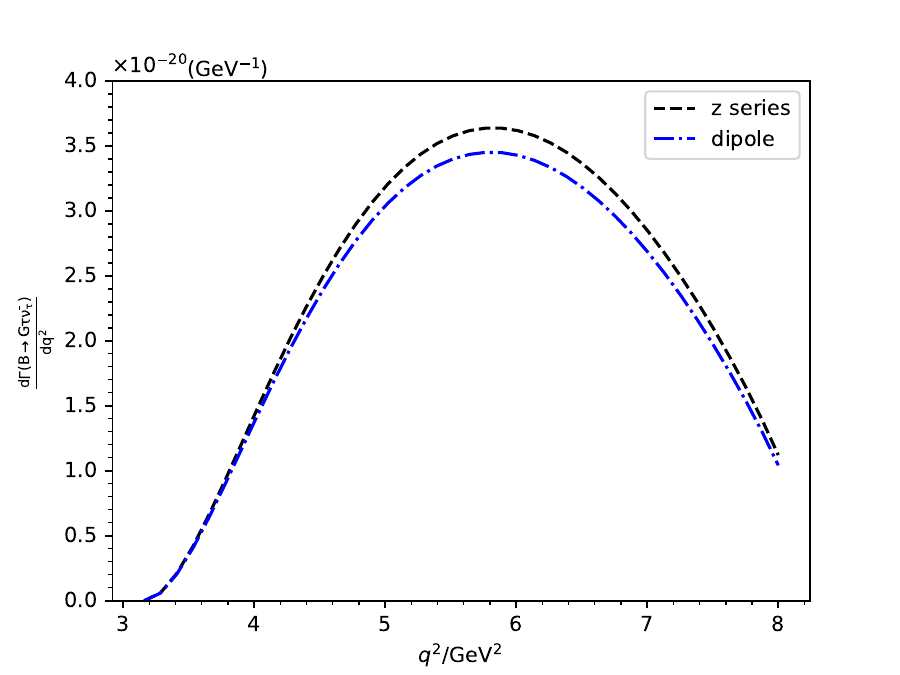}
\end{center}
\caption{The differential decay widths as functions of $q^2$ where the upper panel denotes the decay $B\rightarrow G l\Bar{\nu}(l=e,\mu)$ and the lower panel denotes the decay $B\rightarrow G\tau\Bar{\nu_\tau}$.} \label{diagram:q2decaywidth}
\end{figure}

The partial decay width of $B\rightarrow Gl\Bar{\nu}$\cite{Wang:2009rc}  is given by
\begin{eqnarray}
\frac{d\Gamma(B\rightarrow Gl\Bar{\nu})}{dq^2}=\frac{\sqrt{\lambda}G_F^2\left|V_{ub}\right|^2}{384\pi^3m_B^3q^2}\left(\frac{q^2-m_l^2}{q^2}\right)^2 \times\nonumber\\
\left[(m_l^2+2q^2)\lambda F_1^2(q^2)+3m_l^2(m_B^2-m_G^2)^2F_0^2(q^2)\right],
\end{eqnarray}
where $\lambda=(m_B^2-q^2-m_G^2)^2-4m_G^2 q^2$ and $m_l(l=e,\mu)$ is the lepton mass. The CKM matrix element is taken as $\left|V_{ub}\right|=(3.70\pm0.10\pm0.13)\times10^{-3}$ and Fermi constant is taken as $G_F=1.166\times 10^{-5}\rm{GeV}^{-2}$~\cite{ParticleDataGroup:2024cfk}. Different decay width  $d\Gamma/dq^2$ for these decay modes is shown in Fig.~\ref{diagram:q2decaywidth}.

\begin{table}
\caption{The decay width (in units of $10^{-19}\rm{GeV}$) of $B\rightarrow Gl\Bar{\nu}(l=e,\mu)$ and $B\rightarrow G\tau\Bar{\nu_\tau}$ are obtained with PQCD factorization. The errors are due to the uncertainties in the values of $\omega_B=(0.40\pm0.05)\mbox{GeV}$, $f_B=(0.19\pm0.02)\rm{GeV}$ and $\left|V_{ub}\right|=(3.70\pm0.16)\times10^{-3}$~~\cite{ParticleDataGroup:2024cfk}.}\label{tab:gamma}
\renewcommand{\arraystretch}{2.0}
\setlength{\tabcolsep}{2.5mm}
\begin{tabular}{c|c}
\hline\hline & PQCD factorization 
 \\ \hline
$ \Gamma\left( B\rightarrow Gl\Bar{\nu} \right)_{dip}$  &$7.97^{+3.04 +1.62 +0.74 }_{-2.09 -1.69 -0.68}$
\\ \hline
$ \Gamma\left( B\rightarrow Gl\Bar{\nu} \right)_{z}$  &$7.63^{+3.24 +2.02 +0.68 }_{-1.89 -1.35 -0.68}$
\\ \hline
$\Gamma(B\rightarrow G\tau\Bar{\nu_\tau})_{dip}$  &$1.08^{+0.40 +0.20 +0.14 }_{-0.27 -0.20 -0.07}$
\\ \hline
$\Gamma(B\rightarrow G\tau\Bar{\nu_\tau})_{z}$  &$1.08^{+0.47 +0.27 +0.07 }_{-0.27 -0.20 -0.07}$
\\ 
\hline\hline
\end{tabular}
\end{table}

Using the lifetime of $B$ meson $\tau_B=1.638\times 10^{-12}\rm{s}$ ~\cite{ParticleDataGroup:2024cfk}, we can calculate the total decay width of B meson: 
\begin{eqnarray}
\Gamma (B)=\hbar/\tau_B=4.02\times10^{-13}\rm{GeV}.
\end{eqnarray}

By integrating the partial decay width with $q^2$ we obtain the decay width
\begin{eqnarray}
\Gamma (\bar{B}\rightarrow Gl\Bar{\nu})=\int_{m_l^2}^{(m_B-m_G)^2}\frac{d\Gamma(B\rightarrow Gl\Bar{\nu})}{dq^2}dq^2.
\end{eqnarray}
Numerical results are shown in table~\ref{tab:gamma}.
By adding these uncertainties in quadrature, we obtain the total theoretical uncertainty for these branching ratios, which includes the uncertainties from the form factors and CKM matrix elements. The branching  ratios of $B\rightarrow Gl\Bar{\nu}(l=e,\mu)$ and $B\rightarrow G\tau\Bar{\nu_\tau}$ calculated using the PQCD approach are as follows:
\begin{eqnarray}
&&\mathcal{B}\left( B\rightarrow Gl\Bar{\nu} \right)_{dip}=1.96^{+0.88}_{-0.68}\times 10^{-6}, \\
&&\mathcal{B}\left(B\rightarrow Gl\Bar{\nu} \right)_{z}=1.89^{+0.95}_{-0.61}\times 10^{-6},\\
&&\mathcal{B}(B\rightarrow G\tau\Bar{\nu_\tau})_{dip}=0.27^{+0.11}_{-0.09}\times 10^{-6},\\
&&\mathcal{B}(B\rightarrow G\tau\Bar{\nu_\tau})_{z}=0.26^{+0.14}_{-0.08}\times 10^{-6}.
\end{eqnarray}
Results from different parametrizations are consistent with each other.

As shown in Fig.~\ref{diagram:q2decaywidth} for the differential decay widths, we observe that the dipole parametrization results in a smoother and more gradual increase in the decay width with increasing $q^2$. In contrast, the $z$-series parametrization leads to a more abrupt rise, particularly at lower $q^2$ values. This difference is attributed to the different ways in which the form factors are extrapolated to higher $q^2$ regions beyond the PQCD reliable region. The decay widths as functions of $q^2$ obtained from the PQCD factorization method are consistently lower than those from the collinear factorization method. This discrepancy arises from the inclusion of Sudakov factors and threshold resummation effects in the PQCD approach.

\subsection{Nonleptonic Decays}
For the decays $B^{-} \rightarrow \pi^{-}+G$($B^{-} \rightarrow K^{-}+G$), the leading order weak effective Hamiltonian $H_{\rm eff}$
can be expressed as follows~\cite{Buchalla:1995vs}: 
\begin{equation}
H_{\mathrm{eff}}=\frac{G_F}{\sqrt{2}}V_{u b} V_{u q}^*\left[C_1(\mu) O_1^u(\mu)+C_2(\mu) O_2^u(\mu)\right]+\rm H.\rm c.,
\label{eq:heff} 
\end{equation} 
where the light quark $q$ can be either $d$ or $s$. The $C_i(\mu)$ are the Wilson coefficients evaluated at the renormalization scale $\mu$. The local four-quark operators $O_i$ are categorized as tree operators as follow:
\begin{eqnarray}
&&O_1^{u}\, =\,
(\bar{u}_\alpha b_\beta)_{V-A}(\bar{q}_\beta u_\alpha)_{V-A}\;,
\nonumber \\
&&O_2^{u}\, =\, (\bar{u}_\alpha b_\alpha)_{V-A}(\bar{q}_\beta d_\beta)_{V-A},
\end{eqnarray}
where $\alpha$ and $\beta$ are the color indices and the notation $(V-A)$ refers to the Lorentz structure $\gamma_{\mu}(1-\gamma_{5})$.

$B^{-} \rightarrow P^{-}+G$ amplitudes can be written as
\begin{eqnarray}
&&\mathcal{M}\left(B^{-} \rightarrow P^{-}G\right)
\nonumber\\
&&=-\frac{1}{\sqrt{2}}i f_P a_{1}G_{F} V_{u b} V_{u q}^{*}F_{0}(m_{P^-}^2)(m_B^2-m_G^2),
\end{eqnarray}
where $P$ denotes $\pi$ and $K$. $f_P$ is the decay constant of the corresponding $P$ meson. Here  the  effective Wilson coefficients is used  as $a_{1}=C_{2}+\frac{C_{1}}{N_C}=1.07$~\cite{Buchalla:1995vs}. 

The decay widths for $B^{-} \rightarrow P^{-} G$ are given as follows
\begin{eqnarray}
\Gamma\left(B^{-} \rightarrow P^{-}G\right)=\frac{|\vec{p}|}{8 \pi m_B^2}\left|\mathcal{M}\left(B^{-} \rightarrow P^{-}G\right)\right|^2,
\end{eqnarray}
with $|\vec{p}|=\sqrt{\frac{(m_B^2-m_G^2+m_{P^-}^2)^2}{4 m_B^2}-m_{P^-}^2}$.

In doing the numerical calculation, we adopt the following values for the input parameters:
\begin{eqnarray}
&& m_{\pi^\pm}=0.140\,{\rm GeV},\; m_{K^\pm}=0.494\,{\rm GeV},\; \nonumber \\
&& f_{\pi}=0.13\, {\rm GeV}, \;\,f_{K}=0.16\,{\rm GeV}, \nonumber \\
&& \left|V_{ud}\right|=0.973,\;\,\left|V_{us}\right|=0.224.
\label{eq.input}
\end{eqnarray}
The adopted values of the meson masses and the CKM matrix elements from PDG~\cite{ParticleDataGroup:2024cfk}. The CKM matrix $\left|V_{ud}\right|$ and $\left|V_{us}\right|$ has been measured with high precision, hence the associated errors can be disregarded, in this work.

The decay widths $B^{\pm} \rightarrow \pi^{\pm}+G$ and $B^{\pm} \rightarrow K^{\pm}+G$ are calculated using PQCD factorization with theoretical total errors from form factor and $\left|V_{ub}\right|$. The values for decay widths in PQCD approach are shown as:
\begin{eqnarray}
\Gamma\left(B^{\pm} \rightarrow \pi^{\pm}\;G\right)_{dip}=2.0^{+1.0}_{-0.7}\times10^{-19}\rm{GeV},\\
\Gamma\left(B^{\pm} \rightarrow \pi^{\pm}\;G\right)_{z}=1.7^{+1.4}_{-0.6}\times10^{-19}\rm{GeV},\\
\Gamma\left(B^{\pm} \rightarrow K^{\pm}\;G\right)_{dip}=0.16^{+0.08}_{-0.06}\times10^{-19}\rm{GeV},\\
\Gamma\left(B^{\pm} \rightarrow K^{\pm}\;G\right)_{z}=0.14^{+0.11}_{-0.05}\times10^{-19}\rm{GeV}.
\end{eqnarray}
The branching ratios calculated using the PQCD approach are as follows:
\begin{eqnarray}
&&\mathcal{B}\left(B^{\pm} \rightarrow \pi^{\pm}\;G\right)_{dip}= 4.9^{+2.4}_{-1.8}\times 10^{-7},\\
&&\mathcal{B}\left(B^{\pm} \rightarrow \pi^{\pm}\;G\right)_{z}= 4.3^{+3.4}_{-1.5}
\times 10^{-7},\\
&&\mathcal{B}\left(B^{\pm} \rightarrow K^{\pm}\;G\right)_{dip}= 0.38^{+0.18}_{-0.15}\times 10^{-7},\\
&&\mathcal{B}\left(B^{\pm} \rightarrow K^{\pm} \; G\right)_{z}= 0.34^{+0.27}_{-0.12}\times 10^{-7}.
\end{eqnarray}


At Belle-II, there will be roughly $5 \times 10^{10}$ events involving $b$ quarks~\cite{Belle-II:2018jsg} and approximately $10^{10}$ events of $B$ hadrons produced. These figures translate to tens to hundreds of thousands of individual events. Consequently, the detection of this decay channel and the subsequent investigation of the $X(2370)$ particle’s properties are highly probable.

\section{Summary}

The BESIII experiment has recently conducted an insightful analysis of the $X(2370)$ particle, focusing on its mass and spin parity characteristics. Results obtained from this study suggest a compelling alignment with the anticipated features of the lightest pseudoscalar glueball. To delve deeper into the intricate nature of the $X(2370)$ particle,  we have explored investigations centered on heavy meson decays. By assuming the identity of $X(2370)$ as a pseudoscalar glueball, we have computed form factors for transitions such as $B\to X(2370)$ using a factorization approach. Notably, the estimated branching fractions for $B$ semileptonic decays leading to $X(2370)$ were calculated to be on the order of $10^{-6}$-$10^{-8}$, indicating the potential for the detection of $X(2370)$ decays into the final state $K_S^0K_S^0\eta'$.

Though higher order QCD corrections are not taken into account, these findings pave the way for future experimental endeavors poised to expand our current understanding of glueball physics. The prospects of unraveling the mysteries surrounding the $X(2370)$ particle hold great promise for advancing scientific discourse in this field. By continuing to probe the properties and behaviors of $X(2370)$ through experimental exploration, researchers aim to shed further light on the enigmatic world of glueball physics, ultimately contributing valuable insights to the ongoing discussions and debates surrounding this intriguing particle.

\section*{Acknowledgement}
This work is supported in part by Natural Science Foundation of China under grant No. 12125503, 12335003 and 12375069. 

\begin{appendix}

\section{Hard kernels}

The hard functions $h(x_1,x_2,b_1,b_2)$ come form the Fourier transform and can be written as
\begin{align}
h(x_1,&x_2,b_1,b_2)=K_0(\sqrt{\rho \bar x_2 x_1}
m_{B}b_2)\nonumber\\
& \times\left[\theta(b_1-b_2)I_0(\sqrt
{\rho x_1}m_{B}b_2)K_0(\sqrt
{\rho x_1} m_{B}b_1)\right.  \nonumber\\
& + \left.\theta(b_2-b_1)I_0(\sqrt {\rho x_1}m_{B}b_1)K_0(\sqrt
{\rho x_1}m_{B}b_2)\right], 
\end{align}
where  $K_0$, $I_0$ are modified Bessel functions.
The Sudakov factors are given by:
\begin{eqnarray}
S_B(t)&=&s\left (x_1\frac{m_B}{\sqrt{2}},b_1 \right ) \nonumber\\
&&+\frac{5}{3}\int_{1/b_1}^{t}
\frac{d\bar{\mu}}{\bar{\mu}}\gamma_q\left (\alpha_s(\bar{\mu}) \right ),\\
S_G(t)&=&s\left (x_2m_B,b_2 \right )
+s\left ((1-x_2)m_B,b_2 \right ) \nonumber\\
&& 
+2\int_{1/b_2}^{t}\frac{d\bar{\mu}}{\bar{\mu}}\gamma_q\left (\alpha_s(\bar{\mu}) \right ),
\end{eqnarray}
with anomalous dimensions $\gamma_q=-\alpha_s/\pi$.
The Sudakov function $s(Q,b)$ can be presented in the following general form
\begin{eqnarray}
s(Q,b)&=&\frac{A^{(1)}}{2\beta_1}\hat{q}\ln\left (\frac{\hat{q}}{\hat{b}}\right)
-\frac{A^{(1)}}{2\beta_1}(\hat{q}-\hat{b})
+\frac{A^{(2)}}{4\beta_1^2}\left (\frac{\hat{q}}{\hat{b}}-1 \right )  \nonumber \\
&&-\left [\frac{A^{(2)}}{4\beta_1^2}-\frac{A^{(1)}}{4\beta_1}
\ln\left (\frac{e^{2\gamma_E}-1}{2}\right )\right ]
\ln\left ({\frac{\hat{q}}{\hat{b}}}\right ) \nonumber \\
&& +\frac{A^{(1)}\beta_2}{4\beta_1^3}\hat{q}\left [\frac{\ln(2\hat{q})+1}{\hat{q}}
-\frac{\ln(2\hat{b})+1}{\hat{b}} \right ] \nonumber \\
&&+\frac{A^{(1)}\beta_2}{8\beta_1^3} \left [ \ln^2(2\hat{q})-\ln^2(2\hat{b})\right ].\label{eq:suda-form}
\end{eqnarray}
The variables $\hat{q}$ and $\hat{b}$ are defined as follows: 
\begin{eqnarray}
\hat{q}=\ln[Q/(\sqrt{2}\Lambda)],\quad \hat{b}=\ln[1/(b\Lambda)] ,
\end{eqnarray}
and the coefficients $A^{(i)}$ and $\beta_i$ are
\begin{eqnarray}
\beta_1&=&\frac{33-2n_f}{12}, \quad \beta_2=\frac{153-19n_f}{24},\quad
A^{(1)}=\frac{4}{3}, \notag \\
\quad A^{(2)}&=&\frac{67}{9}-\frac{\pi^2}{3}-\frac{10n_f}{27}
+\frac{8}{3}\beta_1\ln(e^{\gamma_E}/2).
\end{eqnarray}
Indeed, only the leading order $s(Q,b)$ needs to be considered in this work.
\end{appendix}
\hspace{2cm}


\begin{thebibliography}{1}

\bibitem{Klempt:2007cp}
E.~Klempt and A.~Zaitsev,
Phys. Rept. \textbf{454}, 1-202 (2007)
doi:10.1016/j.physrep.2007.07.006
[arXiv:0708.4016 [hep-ph]].

\bibitem{Crede:2008vw}
V.~Crede and C.~A.~Meyer,
Prog. Part. Nucl. Phys. \textbf{63}, 74-116 (2009)
doi:10.1016/j.ppnp.2009.03.001
[arXiv:0812.0600 [hep-ex]].

\bibitem{BESIII:2023wfi}
M.~Ablikim \textit{et al.} [BESIII],
Phys. Rev. Lett. \textbf{132}, no.18, 181901 (2024)
doi:10.1103/PhysRevLett.132.181901
[arXiv:2312.05324 [hep-ex]].

\bibitem{Liu:2010tr}
J.~F.~Liu \textit{et al.} [BES],
Phys. Rev. D \textbf{82}, 074026 (2010)
doi:10.1103/PhysRevD.82.074026
[arXiv:1008.0246 [hep-ph]].

\bibitem{Gui:2019dtm}
L.~C.~Gui, J.~M.~Dong, Y.~Chen and Y.~B.~Yang,
Phys. Rev. D \textbf{100}, no.5, 054511 (2019)
doi:10.1103/PhysRevD.100.054511
[arXiv:1906.03666 [hep-lat]].

\bibitem{Bali:1993fb}
G.~S.~Bali \textit{et al.} [UKQCD],
Phys. Lett. B \textbf{309}, 378-384 (1993)
doi:10.1016/0370-2693(93)90948-H
[arXiv:hep-lat/9304012 [hep-lat]].

\bibitem{Morningstar:1999rf}
C.~J.~Morningstar and M.~J.~Peardon,
Phys. Rev. D \textbf{60}, 034509 (1999)
doi:10.1103/PhysRevD.60.034509
[arXiv:hep-lat/9901004 [hep-lat]].

\bibitem{Chen:2005mg}
Y.~Chen, A.~Alexandru, S.~J.~Dong, T.~Draper, I.~Horvath, F.~X.~Lee, K.~F.~Liu, N.~Mathur, C.~Morningstar and M.~Peardon, \textit{et al.}
Phys. Rev. D \textbf{73}, 014516 (2006)
doi:10.1103/PhysRevD.73.014516
[arXiv:hep-lat/0510074 [hep-lat]].

\bibitem{Gregory:2012hu}
E.~Gregory, A.~Irving, B.~Lucini, C.~McNeile, A.~Rago, C.~Richards and E.~Rinaldi,
JHEP \textbf{10}, 170 (2012)
doi:10.1007/JHEP10(2012)170
[arXiv:1208.1858 [hep-lat]].

\bibitem{Gui:2012gx}
L.~C.~Gui \textit{et al.} [CLQCD],
Phys. Rev. Lett. \textbf{110}, no.2, 021601 (2013)
doi:10.1103/PhysRevLett.110.021601
[arXiv:1206.0125 [hep-lat]].

\bibitem{Zou:2024ksc}
J.~Zou, L.~C.~Gui, Y.~Chen, W.~Qin, J.~Liang, X.~Jiang and Y.~Yang,
Sci. China Phys. Mech. Astron. \textbf{67}, no.11, 111012 (2024)
doi:10.1007/s11433-024-2451-5
[arXiv:2404.01564 [hep-lat]].

\bibitem{Yang:2013xba}
Y.~B.~Yang \textit{et al.} [CLQCD],
Phys. Rev. Lett. \textbf{111}, no.9, 091601 (2013)
doi:10.1103/PhysRevLett.111.091601
[arXiv:1304.3807 [hep-lat]].

\bibitem{Jiang:2022ffl}
X.~Jiang, W.~Sun, F.~Chen, Y.~Chen, M.~Gong, Z.~Liu and R.~Zhang,
Phys. Rev. D \textbf{107}, no.9, 094510 (2023)
doi:10.1103/PhysRevD.107.094510
[arXiv:2205.12541 [hep-lat]].

\bibitem{Yu:2011ta}
J.~S.~Yu, Z.~F.~Sun, X.~Liu and Q.~Zhao,
Phys. Rev. D \textbf{83}, 114007 (2011)
doi:10.1103/PhysRevD.83.114007
[arXiv:1104.3064 [hep-ph]].

\bibitem{Deng:2012wi}
C.~Deng, J.~Ping, Y.~Yang and F.~Wang,
Phys. Rev. D \textbf{86}, 014008 (2012)
doi:10.1103/PhysRevD.86.014008
[arXiv:1202.4167 [hep-ph]].

\bibitem{She:2024ewy}
Z.~L.~She, A.~K.~Lei, W.~C.~Zhang, Y.~L.~Yan, D.~M.~Zhou, H.~Zheng and B.~H.~Sa,
[arXiv:2407.07661 [hep-ph]].

\bibitem{Cao:2024mfn}
J.~Cao, Z.~L.~She, J.~P.~Zhang, J.~H.~Shi, Z.~Y.~Qin, W.~C.~Zhang, H.~Zheng, A.~K.~Lei, D.~M.~Zhou and Y.~L.~Yan, \textit{et al.}
Phys. Rev. D \textbf{110}, no.5, 054046 (2024)
doi:10.1103/PhysRevD.110.054046
[arXiv:2408.04130 [hep-ph]].

\bibitem{Li:2024fko}
H.~n.~Li,
Chin. Phys. Lett. \textbf{41}, no.10, 101101 (2024)
doi:10.1088/0256-307X/41/10/101101
[arXiv:2408.06738 [hep-ph]].

\bibitem{He:2006qk}
X.~G.~He and T.~C.~Yuan,
[arXiv:hep-ph/0612108 [hep-ph]].

\bibitem{Charng:2006zj}
Y.~Y.~Charng, T.~Kurimoto and H.~n.~Li,
Phys. Rev. D \textbf{74}, 074024 (2006)
[erratum: Phys. Rev. D \textbf{78}, 059901 (2008)]
doi:10.1103/PhysRevD.78.059901
[arXiv:hep-ph/0609165 [hep-ph]].

\bibitem{Wang:2009rc}
W.~Wang, Y.~L.~Shen and C.~D.~Lu,
J. Phys. G \textbf{37}, 085006 (2010)
doi:10.1088/0954-3899/37/8/085006
[arXiv:0908.2216 [hep-ph]].

\bibitem{Lu:2013jj}
C.~D.~L\"u, U.~G.~Meissner, W.~Wang and Q.~Zhao,
Eur. Phys. J. A \textbf{49}, 58 (2013)
doi:10.1140/epja/i2013-13058-y
[arXiv:1301.0225 [hep-ph]].

\bibitem{Huang:2021ots}
F.~Huang and Q.~A.~Zhang,
Eur. Phys. J. C \textbf{82}, no.1, 11 (2022)
doi:10.1140/epjc/s10052-021-09779-1
[arXiv:2108.06110 [hep-ph]].

\bibitem{Wang:2017hxe}
C.~Wang, Q.~A.~Zhang, Y.~Li and C.~D.~Lu,
Eur. Phys. J. C \textbf{77}, no.5, 333 (2017)
doi:10.1140/epjc/s10052-017-4889-3
[arXiv:1701.01300 [hep-ph]].

\bibitem{Zhou:2016jkv}
S.~H.~Zhou, Q.~A.~Zhang, W.~R.~Lyu and C.~D.~L\"u,
Eur. Phys. J. C \textbf{77}, no.2, 125 (2017)
doi:10.1140/epjc/s10052-017-4685-0
[arXiv:1608.02819 [hep-ph]].

\bibitem{Keum:2000wi}
Y.~Y.~Keum, H.~N.~Li and A.~I.~Sanda,
Phys. Rev. D \textbf{63}, 054008 (2001)
doi:10.1103/PhysRevD.63.054008
[arXiv:hep-ph/0004173 [hep-ph]].

\bibitem{Keum:2000ph}
Y.~Y.~Keum, H.~n.~Li and A.~I.~Sanda,
Phys. Lett. B \textbf{504}, 6-14 (2001)
doi:10.1016/S0370-2693(01)00247-7
[arXiv:hep-ph/0004004 [hep-ph]].

\bibitem{Lu:2000em}
C.~D.~Lu, K.~Ukai and M.~Z.~Yang,
Phys. Rev. D \textbf{63}, 074009 (2001)
doi:10.1103/PhysRevD.63.074009
[arXiv:hep-ph/0004213 [hep-ph]].

\bibitem{Belle-II:2018jsg}
E.~Kou \textit{et al.} [Belle-II],
PTEP \textbf{2019}, no.12, 123C01 (2019)
[erratum: PTEP \textbf{2020}, no.2, 029201 (2020)]
doi:10.1093/ptep/ptz106
[arXiv:1808.10567 [hep-ex]].

\bibitem{Grozin:1996pq}
A.~G.~Grozin and M.~Neubert,
Phys. Rev. D \textbf{55}, 272-290 (1997)
doi:10.1103/PhysRevD.55.272
[arXiv:hep-ph/9607366 [hep-ph]].

\bibitem{Kawamura:2001jm}
H.~Kawamura, J.~Kodaira, C.~F.~Qiao and K.~Tanaka,
Phys. Lett. B \textbf{523}, 111 (2001)
[erratum: Phys. Lett. B \textbf{536}, 344-344 (2002)]
doi:10.1016/S0370-2693(01)01299-0
[arXiv:hep-ph/0109181 [hep-ph]].

\bibitem{Ball:2004ye}
P.~Ball and R.~Zwicky,
Phys. Rev. D \textbf{71}, 014015 (2005)
doi:10.1103/PhysRevD.71.014015
[arXiv:hep-ph/0406232 [hep-ph]].

\bibitem{Hu:2012cp}
H.~C.~Hu and H.~n.~Li,
Phys. Lett. B \textbf{718}, 1351-1357 (2013)
doi:10.1016/j.physletb.2012.12.006
[arXiv:1204.6708 [hep-ph]].

\bibitem{Li:2002uya}
H.~n.~Li,
Nucl. Phys. B Proc. Suppl. \textbf{111}, 69-74 (2002)
doi:10.1016/S0920-5632(02)01686-9

\bibitem{ParticleDataGroup:2024cfk}
S.~Navas \textit{et al.} [Particle Data Group],
Phys. Rev. D \textbf{110}, no.3, 030001 (2024)
doi:10.1103/PhysRevD.110.030001

\bibitem{Han:2024min}
X.~Y.~Han, J.~Hua, X.~Ji, C.~D.~L\"u, W.~Wang, J.~Xu, Q.~A.~Zhang and S.~Zhao,
[arXiv:2403.17492 [hep-ph]].

\bibitem{Han:2024yun}
X.~Y.~Han, J.~Hua, X.~Ji, C.~D.~L\"u, A.~Sch\"afer, Y.~Su, W.~Wang, J.~Xu, Y.~Yang and J.~H.~Zhang, \textit{et al.}
[arXiv:2410.18654 [hep-lat]].

\bibitem{Wang:2024wwa}
W.~Wang, J.~Xu, Q.~A.~Zhang and S.~Zhao,
[arXiv:2411.07101 [hep-ph]].

\bibitem{LatticeParton:2020uhz}
Q.~A.~Zhang \textit{et al.} [Lattice Parton],
Phys. Rev. Lett. \textbf{125}, no.19, 192001 (2020)
doi:10.22323/1.396.0477
[arXiv:2005.14572 [hep-lat]].

\bibitem{LatticePartonLPC:2022eev}
M.~H.~Chu \textit{et al.} [Lattice Parton (LPC)],
Phys. Rev. D \textbf{106}, no.3, 034509 (2022)
doi:10.1103/PhysRevD.106.034509
[arXiv:2204.00200 [hep-lat]].

\bibitem{LatticeParton:2023xdl}
M.~H.~Chu \textit{et al.} [Lattice Parton],
Phys. Rev. D \textbf{109}, no.9, L091503 (2024)
doi:10.1103/PhysRevD.109.L091503
[arXiv:2302.09961 [hep-lat]].

\bibitem{LatticePartonLPC:2023pdv}
M.~H.~Chu \textit{et al.} [Lattice Parton (LPC)],
JHEP \textbf{08}, 172 (2023)
doi:10.1007/JHEP08(2023)172
[arXiv:2306.06488 [hep-lat]].

\bibitem{Chu:2024vkn}
M.~H.~Chu, H.~Bai, J.~Hua, J.~Liang, X.~Ji, A.~Schafer, Y.~Su, W.~Wang, Y.~B.~Yang and J.~Zeng, \textit{et al.}
[arXiv:2411.12554 [hep-lat]].

\bibitem{LatticeParton:2022zqc}
J.~Hua \textit{et al.} [Lattice Parton],
Phys. Rev. Lett. \textbf{129}, no.13, 132001 (2022)
doi:10.1103/PhysRevLett.129.132001
[arXiv:2201.09173 [hep-lat]].

\bibitem{Liu:2018tox}
Y.~S.~Liu, W.~Wang, J.~Xu, Q.~A.~Zhang, S.~Zhao and Y.~Zhao,
Phys. Rev. D \textbf{99}, no.9, 094036 (2019)
doi:10.1103/PhysRevD.99.094036
[arXiv:1810.10879 [hep-ph]].

\bibitem{Liu:2019urm}
Y.~S.~Liu, W.~Wang, J.~Xu, Q.~A.~Zhang, J.~H.~Zhang, S.~Zhao and Y.~Zhao,
Phys. Rev. D \textbf{100}, no.3, 034006 (2019)
doi:10.1103/PhysRevD.100.034006
[arXiv:1902.00307 [hep-ph]].

\bibitem{LatticeParton:2018gjr}
Y.~S.~Liu \textit{et al.} [Lattice Parton],
Phys. Rev. D \textbf{101}, no.3, 034020 (2020)
doi:10.1103/PhysRevD.101.034020
[arXiv:1807.06566 [hep-lat]].

\bibitem{Hua:2020gnw}
J.~Hua \textit{et al.} [Lattice Parton],
Phys. Rev. Lett. \textbf{127}, no.6, 062002 (2021)
doi:10.1103/PhysRevLett.127.062002
[arXiv:2011.09788 [hep-lat]].

\bibitem{Xu:2018mpf}
J.~Xu, Q.~A.~Zhang and S.~Zhao,
Phys. Rev. D \textbf{97}, no.11, 114026 (2018)
doi:10.1103/PhysRevD.97.114026
[arXiv:1804.01042 [hep-ph]].

\bibitem{LatticePartonCollaborationLPC:2022myp}
J.~C.~He \textit{et al.} [Lattice Parton Collaboration (LPC)],
Phys. Rev. D \textbf{109}, no.11, 114513 (2024)
doi:10.1103/PhysRevD.109.114513
[arXiv:2211.02340 [hep-lat]].

\bibitem{He:2002hr}
X.~G.~He, H.~Y.~Jin and J.~P.~Ma,
Phys. Rev. D \textbf{66}, 074015 (2002)
doi:10.1103/PhysRevD.66.074015
[arXiv:hep-ph/0203191 [hep-ph]].


\bibitem{Ali:2003kg}
A.~Ali and A.~Y.~Parkhomenko,
Eur. Phys. J. C \textbf{30} (2003), 367-380
doi:10.1140/epjc/s2003-01302-6
[arXiv:hep-ph/0307092 [hep-ph]].


\bibitem{Li:2001ay}
H.~n.~Li,
Phys. Rev. D \textbf{66}, 094010 (2002)
doi:10.1103/PhysRevD.66.094010
[arXiv:hep-ph/0102013 [hep-ph]].

\bibitem{Lim:1998uc}
J.~L.~Lim and H.~n.~Li,
Chin. J. Phys. \textbf{38}, 801-813 (2000)
[arXiv:hep-ph/9807437 [hep-ph]].

\bibitem{Bourrely:2008za}
C.~Bourrely, I.~Caprini and L.~Lellouch,
Phys. Rev. D \textbf{79}, 013008 (2009)
[erratum: Phys. Rev. D \textbf{82}, 099902 (2010)]
doi:10.1103/PhysRevD.82.099902
[arXiv:0807.2722 [hep-ph]].

\bibitem{Zhang:2021oja}
Q.~A.~Zhang, J.~Hua, F.~Huang, R.~Li, Y.~Li, C.~L\"u, C.~D.~Lu, P.~Sun, W.~Sun and W.~Wang, \textit{et al.}
Chin. Phys. C \textbf{46}, no.1, 011002 (2022)
doi:10.1088/1674-1137/ac2b12
[arXiv:2103.07064 [hep-lat]].

\bibitem{Buchalla:1995vs}
G.~Buchalla, A.~J.~Buras and M.~E.~Lautenbacher,
Rev. Mod. Phys. \textbf{68}, 1125-1144 (1996)
doi:10.1103/RevModPhys.68.1125
[arXiv:hep-ph/9512380 [hep-ph]].

\end{thebibliography}
\end{document}